\documentclass[twocolumn,showpacs,amsmath,amssymb,floatfix]{revtex4}

\usepackage{graphicx}
\begin{document}

\title{  Spin, charge and  single-particle spectral functions of the one-dimensional
         quarter filled Holstein model. }
\author{ F. F. Assaad }

\affiliation{ Institut f\"ur theoretische Physik und Astrophysik,
Universit\"at W\"urzburg, Am Hubland, D-97074 W\"urzburg, Germany }

\begin{abstract}
We use a recently developed extension of the weak coupling diagrammatic determinantal  
quantum Monte Carlo method to investigate the spin, charge and single particle 
spectral functions of the one-dimensional quarter-filled Holstein model  with  
phonon frequency $\omega_0 = 0.1 t$. As a function of the  dimensionless  electron-phonon
coupling  we observe a transition from a Luttinger  to 
a Luther-Emery liquid with dominant $2k_f$ charge fluctuations.  
Emphasis is placed on the temperature dependence of the single particle 
spectral function. At high temperatures and in both phases it is well accounted for within  
a self-consistent Born approximation.   
In the low temperature  Luttinger liquid  phase we observe features which  
compare favorably with a bosonization approach retaining only forward scattering.   
In the Luther-Emery phase, the  spectral function at low temperatures  shows a quasiparticle gap 
which matches half the spin gap whereas at temperatures above which 
this  quasiparticle gap closes,  characteristic features of the Luttinger liquid model are apparent. 
Our  results are based on lattice simulations on chains up to  L=20 for two-particle properties 
and on  CDMFT calculations with clusters up to 12 sites for the  single-particle spectral function.  
\end{abstract}


\maketitle

\section { Introduction }
       Including phonon degrees of freedom in model calculations of correlated electron systems is
challenging but necessary for the  understanding of many experiments. One 
can mention the quasi one-dimensional organics  TTF-TCNQ where  photoemission experiments 
are carried out down to  60 K just above the Peierls transition \cite{Claessen03}. 
A detailed   modeling  of this experimental situation is bound to include both 
electronic  correlations \cite{Koch07,Jeckelmann04,Abendschein06,Bulut06} as well as the phonon 
degrees of freedom  \cite{PhysRevB.14.2325}. In two dimensions the electron-phonon 
interactions leads to a delicate interplay of superconductivity and charge density waves 
depending on the  partial nesting properties of the Fermi surface \cite{Neto01,Borisenko07}.
More generally, the ability to efficiently  include bosonic baths  in Quantum Monte Carlo 
(QMC) simulations is a prerequisite for the implementation of extended  dynamical mean-field
theories (EDMFT) where self-consistency both at the  two particle (bosonic baths) and single particle 
levels is required \cite{Si96,Smith00}.
  
	The aim of this article is to test on  the basis of a  non-trivial model   
a recently proposed generalization 
of  the weak coupling 
diagrammatic determinantal QMC algorithm to include phonon degrees of freedom
\cite{Rubtsov05,Assaad07}. The approach relies on integrating out the phonon degrees of 
freedom at the expense of a retarded  interaction and  then to expand around the 
non-interacting point.  Classes of diagrams  at a given expansion order can be expressed in 
terms of a determinant, the entries of the matrix being the non-interaction Green function. 
The summation over those classes of diagrams is  carried out  with stochastic methods.  
Since the  algorithm action based, the CPU time scales as 
$(\beta L)^3$   ( $\beta$ is the inverse temperature and $L$ the number of lattice sites ) and 
is easily embeded in  dynamical mean-field self-consistency loops.       
To obtain a full account of the physics, we have carried out lattice simulations on 
lattices up to $L=20$ to extract two particle quantities and cluster dynamical mean-field theory
(CDMFT) \cite{Biroli04} calculations on embedded  clusters up to $L_c = 12$ to 
investigate  the single particle  spectral function.
As a function of the dimensionless electron-phonon coupling and at fixed phonon frequency 
$ \omega_0/t = 0.1$, we interpret our low temperature results in terms of a  transition 
from a Luttinger  liquid with gapless  spin and charge modes to a Luther-Emery liquid 
with gapful  spin and gapless charge modes \cite{Voit98}.  
This Peierls phase has dominant  $2k_f$ charge  density wave (CDW)  correlations.  
We have placed emphasis on the temperature dependence of 
the single-particle spectral function in both phases.  At {\it high}  temperatures and in both 
phases the QMC data compares favorably with a self-consistent Born approximation
\cite{Engelsberg63}. The low 
temperature properties in the Luttinger liquid phase compare favorably to a 
bosonization approach retaining only forward scattering \cite{Meden94} 
whereas in the Luther-Emery  phase  a quasiparticle gap matching half the spin gap is apparent. 
The temperature dependence of the single particle spectral function in the Luther-Emery phase 
is particularly rich; at temperature scales where the quasiparticle gap  closes,  
features of the Luttinger liquid model are apparent.       

	The article is organized as follows. In the next section we introduce the model, and 
briefly review our implementation of the  CDMFT.  We refer the reader to Ref. \cite{Assaad07} for 
a detailed description of  the QMC method. In Section  \ref{Numerical_results} we 
present our numerical results for two-particle and single particle  correlation functions 
across the  Peierls transition. 
For  completeness sake, two appendices summarize the 
self-consistent Born approximation \cite{Engelsberg63} and elementary aspects of the 
the Luttinger  model appropriate  for the description of the low-energy excitations of 
the  Luttinger liquid  phase \cite{Meden94}. 
 
\section { Model and Quantum Monte Carlo. }
The one-dimensional Holstein model we consider reads: 
\begin{equation}
\label{Holstein_Ham}
  \hat{H}  =   \sum_{ \pmb{k},\sigma} \epsilon({\pmb k}) \hat{c}^{\dagger}_{ {\pmb k},\sigma}
                       \hat{c}_{{\pmb k}, \sigma} 
           + g \sum_{\pmb i} \hat{Q}_{\pmb i} \hat{n}_{ {\pmb i}}   
                + \sum_{\pmb i} \frac{\hat{P}_{\pmb i}^2}{2M} + \frac{k}{2} \hat{Q}_{\pmb i}^2,
\end{equation}
with tight binding dispersion relation  
$ \epsilon({\pmb k})   = -2t \cos({\pmb k} {\pmb a}) - \mu $.
$ \hat{c}^{\dagger}_{ {\pmb i},\sigma} $ creates an electron in Wannier state centered on lattice site 
$ {\pmb i} $ and with z-component of spin $\sigma $,
$\hat{c}^{\dagger}_{ {\pmb k}, \sigma }  = \frac{1} {\sqrt{L}} \sum_{{\pmb i}} 
e^{ i {\pmb k} \cdot {\pmb i} }
\hat{c}^{\dagger}_{ {\pmb i}, \sigma } $ creates an electron in a 
Bloch state with crystal momentum ${\pmb k}$, 
$ \hat{n}_{ {\pmb i}} = \sum_{\sigma} \hat{c}^{\dagger}_{ {\pmb i}, \sigma }\hat{c}_{ {\pmb i}, \sigma } $ is the on-site particle 
number operator  and $ \hat{Q}_{\pmb i} $ and ${\hat P}_{\pmb i} $ corresponds to the 
ion displacement and momentum.

In  a recent publication \cite{Assaad07}, we have shown how include phonon degrees of freedom 
in the  weak coupling diagrammatic determinantal quantum Monte Carlo (DDQMC) algorithm 
\cite{Rubtsov05}.  The key  ingredient is 
to integrate out the phonon degrees of freedom at the expense of a retarded interaction and 
then to expand around the  non-interacting limit. We refer the reader to Ref. 
\cite{Assaad07} for a detailed description of the algorithm.   

Since  dynamical two particle quantities are  notoriously hard to compute within cluster methods
\cite{Hochkeppel08},  we have used the DDQMC method to simulate the Holstein model on 
lattices up to $L=20$ sites to compute those quantities. 
For the study of the temperature dependence of the
single particle spectral function, we have found it more convenient to  adopt the cluster dynamical 
mean field theory (CDMFT) on embeded cluster sizes up to $L_c = 12$. 

CDMFT  as opposed to  the dynamical cluster approximation (DCA)  is particularly useful to 
tackle our problem.  
It is a real space method  which allows for spontaneous symmetry breaking 
within a predefined unit cell of volume given by the cluster size. To implement the 
method, we  decompose the chain into 
$L_u$, super-cells of length $L_c$.  A site, ${\pmb i}$ in the original lattice then corresponds to 
a  super-cell, ${\pmb R}$, and an orbital index $\nu$ running from $1 \cdots L_c$ such that: 
${\pmb i} = {\pmb R} + {\pmb a}_{\nu} $.    Thereby, the 
volume of the Brillouin zone is reduced by a factor $L_c$ and the quantized  wave vectors 
are given by  ${\pmb K} = \frac{2 \pi }{L_c L_u } n $ with $n \in  [-L_u/2, L_u/2[  $. Within this 
formulation, the self-energy and non-interacting Green function correspond to  
$L_c \times L_c $ matrices, 
${\pmb \Sigma} ( {\pmb K},i \omega_m), {\pmb G}_0 ({\pmb K},i \omega_m)$. 
The CDMFT approximation neglects the  ${\pmb K}$ dependency of the self-energy;
${\pmb \Sigma }({\pmb K},i \omega_m) \equiv {\pmb \Sigma}(i \omega_m) $.  
In analogy to the DMFT approach, one 
can extract the self-energy  by solving on an $L_c$ cluster the model at hand subject to a dynamical 
bath $  {\pmb {\cal  G}}_0(i \omega_m)$ which has to be determined self-consistently. 
To be more precise:
\begin{eqnarray}
\label{CDMFT}
	{ \pmb {\cal G} } (i \omega_m) & = & \frac{1} 
         { { \pmb {\cal G} }^{-1}_0(i \omega_m) - { \pmb \Sigma(i \omega_m)} } 
\nonumber \\
	& = &  \frac{1}{L_u} \sum_{K} \frac{1} { {\pmb G}_0^{-1}(K,i \omega_m) - {\pmb \Sigma}
         (i \omega_m) }.
\end{eqnarray}
The last equality corresponds to self-consistency. Hence, for a given bath Green function matrix
${\pmb {\cal G}}_0 (i \omega_m)$ we use the DDQMC method to  obtain the corresponding  self-energy 
${\pmb \Sigma} (i \omega_m)$ which in turn, owing to Eq. (\ref{CDMFT}), allows us to compute 
a new bath  Green function. This  procedure is repeated till convergence is reached.  
Within the DDQMC  the self-consistency is particularly easy to implement as
it is possible to compute the  Matsubara Green functions directly within the QMC code thus avoiding 
the cumbersome transformation from imaginary time to  Matsubara frequencies. 

Having determined the self-energy, we compute the lattice Green functions, 
$g({\pmb k}, i \omega_m )$  and  ${\pmb k} \in \left[ - \pi,\pi \right]$ with:
\begin{equation}
\begin{gathered}
   g({\pmb k}, i \omega_m ) =  \\
\frac{1}{L_c} \sum_{\mu,\nu = 1}^{L_c} e^{i {\pmb k} 
        \left( {\pmb a}_\mu - {\pmb a}_\nu\right) }  \left[\frac{1}
          { {\pmb G}_0^{-1}( {\pmb K} , i \omega_m) - 
       {\pmb \Sigma} ( i \omega_m) } \right]_{\mu,\nu}. 
\end{gathered}
\end{equation} 
In the above,  $ {\pmb k} = {\pmb K}   + m \frac{2 \pi}{L_c} $ with $ {\pmb K} \in 
\left[ -\frac{\pi}{L_c}, \frac{\pi}{L_c} \right] $. 
We use  CDMFT  solely to extract the single particle spectral function. The required rotation from 
the imaginary to real time axis is accomplished  with a stochastic analytical continuation  scheme
\cite{Sandvik98,Beach04a}.

\section{Numerical Results}
\label{Numerical_results}
In this section, we present our numerical results at quarter filling, $\rho = 0.5$,  phonon 
frequency $\omega_0 = 0.1t$, 
which places us in the adiabatic limit, and  vary the electron-phonon coupling as well 
as the temperature.   We first consider 
spin, charge and  pairing correlations as well as  the optical conductivity  and then  study 
in detail the temperature dependence of the single particle 
spectral function. Two particle quantities are obtained from simulations on an  $L=20$ site lattice.  
To at best study  single particle properties, we have used the CDMFT approximation on cluster sizes 
up to $L_c = 12$.

To characterize the  strength of the electron-phonon interaction, we consider  the  effective
 mass renormalization  as obtained  from the self-energy  diagram shown in Fig. \ref{SCB.fig}. 
For a flat band of width $W$, Eq. (\ref{Meff})  yields:
\begin{equation}
  \frac{m^*}{m} = 1 + \lambda \; \;  {\rm with} \; \; \lambda = \frac{g^2}{2k} \frac{2}{W}
\end{equation}
with  $ \lambda $ the dimensionless electron-phonon coupling.
 
\subsection { Spin and charge static and dynamical structure factors. }

\begin{figure}[ht]
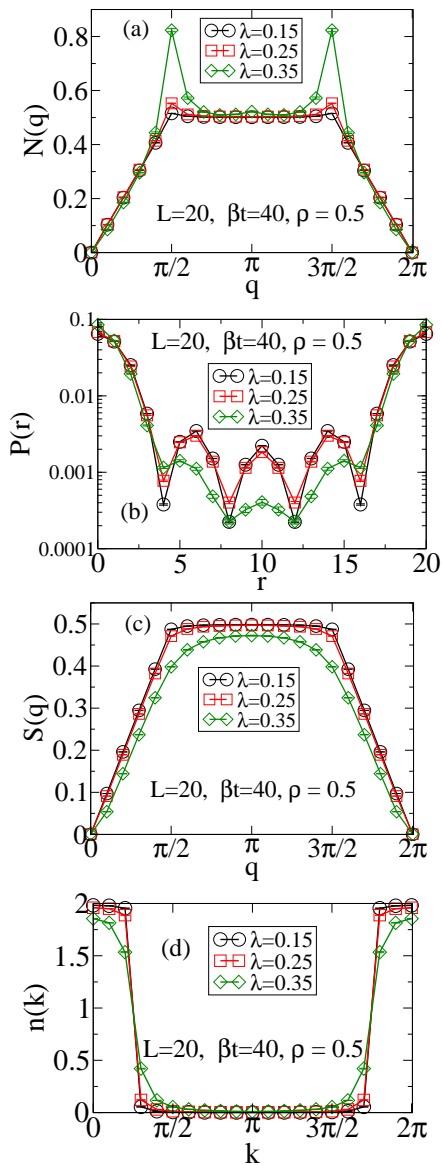

\includegraphics*[width=0.3\textwidth]{DenJ.eps} \\
\vspace{0.1cm}
\includegraphics*[width=0.32\textwidth]{PairJ_R.eps}  \\
\vspace{0.1cm}
\includegraphics*[width=0.3\textwidth]{SpinzJ.eps}  \\
\vspace{0.1cm}
\includegraphics*[width=0.3\textwidth]{XnkJ.eps} \\ 
\caption{ Density (a), pairing (b), spin (c)  correlation functions as well as the single 
particle occupation number (d) as  a function of electron-phonon coupling.  } 
\label{Static.fig}
\end{figure}
Equal time  charge correlation functions,
\begin{equation}
      N({\pmb q}) =  \sum_{\pmb r}  e^{i {\pmb q} {\pmb r}} \left( \langle \hat{n}_{\pmb r}  
      \hat{n}_{0} \rangle -   \langle \hat{n}_{\pmb r} \rangle \langle  \hat{n}_{0} \rangle \right),
\end{equation} 
are plotted in Fig. \ref{Static.fig}a. As a function of growing  electron-phonon coupling, 
the cusp at $2k_F  = \pi/2 $, 
signaling a power-law decay of the correlation function \cite{Assaad91},     evolves towards a clear peak  signaling a dominant $2 k_F $ 
charge modulation  at $\lambda = 0.35$.  Note that  at the largest considered electron-phonon 
coupling,   a cusp  at a higher harmonic, $4 k_F$,  is  equally apparent. 
A simple interpretation of this charge-density wave stems form the Peierls instability.  
For {\it classical}  phonons the inherent $2k_f$ nesting instability of one-dimensional
systems renders the metallic state unstable towards a $2k_f$ lattice deformation at arbitrarily 
small electron phonon coupling. In this mean-field approach the static lattice deformation triggers
the opening of a charge gap. It has been argued and shown numerically \cite{Fehske00} that 
this situation  cannot be  carried over to quantum phonons. In this case, 
quantum fluctuations destroy 
the static lattice deformation and a finite value of  the 
electron-phonon  coupling is required to destabilize the Luttinger liquid.
The linear behavior  of the charge structure factor  at long wavelengths (see Fig. \ref{Static.fig}a)
points to a metallic  state at all considered values of the electron-phonon  interaction 
since it amounts to a powerlaw  decay with {\it modulation} ${\pmb q} =0$ of the real space 
charge correlation function.

Since we have not included  a  Coulomb repulsion in  our model Hamiltonian, 
one expects two electrons of  opposite spins to share the same lattice deformation 
and thereby bind to form bipolarons.   Fig. \ref{Static.fig}b plots the equal time pairing 
correlation functions in the  
on-site s-wave channel, 
\begin{equation}
  P({\pmb r} ) = \langle \hat{\Delta}^{\dagger}_{{\pmb r} } \hat{\Delta}_{{\pmb 0}}  \rangle \; \;  {\rm with} \; \;
  \hat{\Delta}^{\dagger}_{\pmb r}  =  \hat{c}^{\dagger}_{{\pmb r}, \uparrow} \hat{c}^{\dagger}_{{\pmb r},\downarrow}. 
\end{equation}
As apparent, the on-site pairing correlations, $P({\pmb r} =0 )$, grow  as the 
electron-phonon coupling  is enhanced from $\lambda = 0.25 $  to $\lambda = 0.35$.  
This behavior  reflects the formation of bipolarons. On the other hand and in this 
coupling range, the long range pairing correlations are suppressed reflecting a 
tendency towards localization of the bipolarons.

The binding of electrons into spin singlets leads to the suppression of  the $2k_F$ 
spin-spin correlation functions defined by 
\begin{equation}
      S({\pmb q}) =  \sum_{\pmb r}  e^{i {\pmb q} {\pmb r}} \langle \hat{S}_{z,\pmb r} \hat{S}_{z,{\pmb 0}} \rangle
\end{equation}
and plotted in Fig. \ref{Static.fig}c. At $\lambda= 0.35$ both the $q=0$ as well as the $q=2k_F$ 
cusps in the spin  structure factor are smeared out thus lending support to an exponential decay 
of the spin-spin correlation. 

Finally,  the single particle occupation number, 
\begin{equation}
  n({\pmb k})  = \sum_{\sigma} \langle \hat{c}^{\dagger}_{{\pmb k}, \sigma} 
  \hat{c}_{{\pmb k}, \sigma}  \rangle, 
\end{equation}  
is plotted in Fig. \ref{Static.fig}d. As apparent, and on our limited lattice size, $L=20$,  
the jump at  $k_F = \pi/4$ is dramatically suppressed as the electron phonon-interaction  
grows from   $\lambda = 0.25$ to $ \lambda = 0.35$. 

Hence, on the basis of the static quantities, we can conclude that a transition  between  a 
Luttinger liquid metallic phase 
and a spin gaped CDW state  occurs in the region $0.25 < \lambda < 0.35 $.   We now provide 
further support for this picture by examining dynamical two-particle correlation functions.

\begin{figure}[ht]
\includegraphics[width=3.5cm,height=7.5cm]{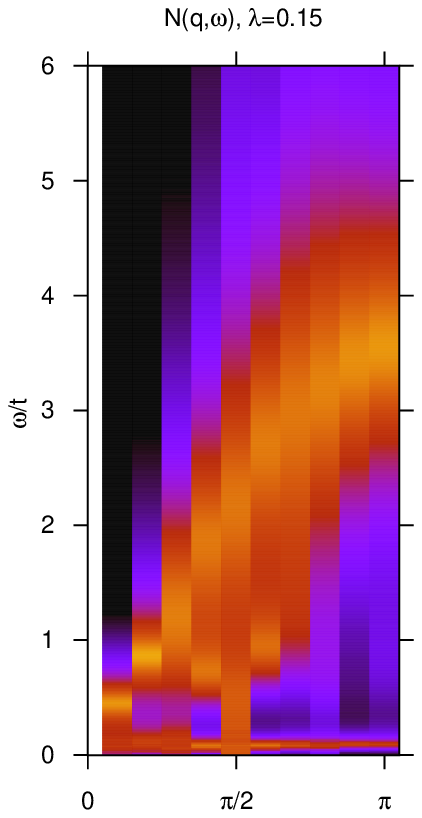} 
\includegraphics[width=4.5cm,height=7.5cm]{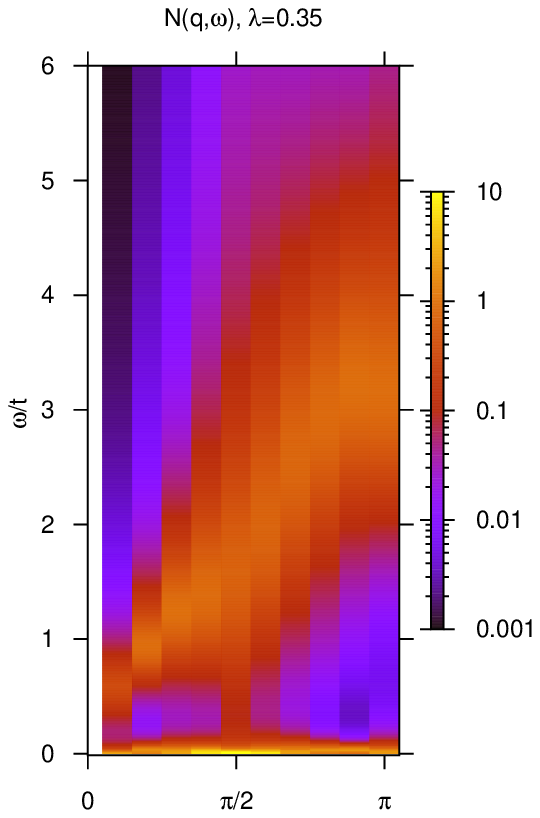}  
\caption{Intensity plots of the dynamical charge structure factor  at $\lambda = 0.15$ 
(left) and $\lambda = 0.35$.  The x-axis corresponds to the momentum $\pmb q$. } 
\label{Dyn_ch.fig}
\end{figure}

In the Lehmann representation, the dynamical charge susceptibility 
is given by:  
\begin{equation}
    N({\pmb q}, \omega) =  \frac{\pi}{Z} \sum_{n,m} e^{-\beta E_m } | \langle n 
    | \hat{n}_{\pmb{q}} | m \rangle|^2 
\delta( E_n -  E_m - \omega )   
\end{equation}
where $ \hat{n}_{\pmb q} = \frac{1}{ \sqrt{N}} \sum_{j} e^{i \pmb{q} {\pmb j} } \hat{n}_{\pmb j} $  
and the  sum rule $ N({\pmb q}) = \frac{1}{\pi} \int d \omega  N({\pmb q},\omega )  $ holds.  
A similar definition holds for the dynamical spin structure factor $S({\pmb q}, \omega)$. 

\begin{figure}[ht]
\includegraphics[width=3.5cm,height=7.5cm]{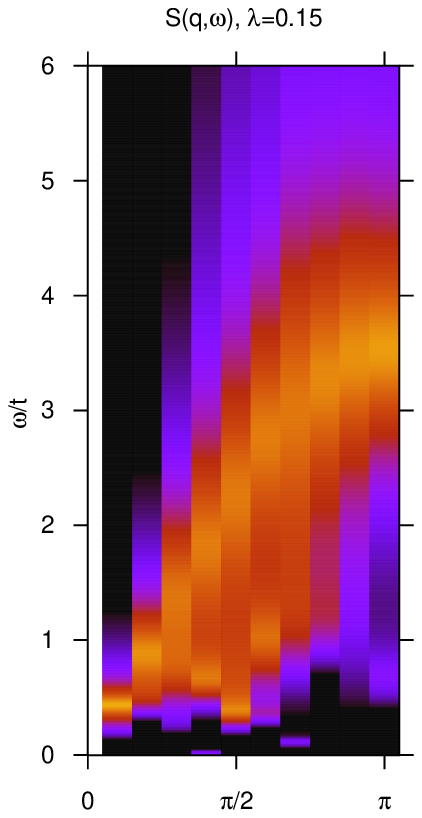} 
\includegraphics[width=4.5cm,height=7.5cm]{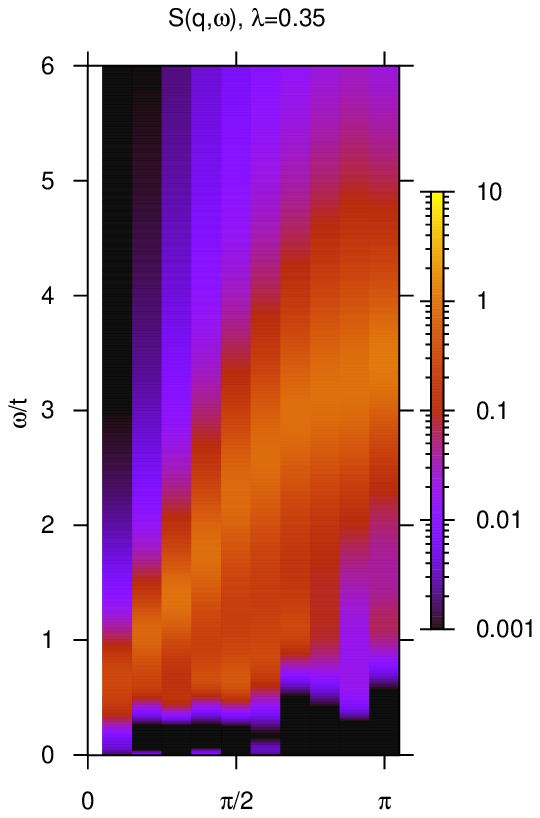}  
\caption{Intensity plots of the dynamical spin structure factor  at $\lambda = 0.15$ 
(left) and $\lambda = 0.35$.  The x-axis corresponds to the momentum $\pmb q$.} 
\label{Dyn_sp.fig}
\end{figure}

In the absence of the electron-phonon coupling, both spin and charge dynamical structure factors 
are identical and correspond to the well know particle-hole 
continuum  with gapless excitations at ${\pmb q} = 0$ and $ {\pmb q} = 2 k_F $.  Note that at
quarter-band filling, $2 k_F = \pi/2 $. As apparent from the Luttinger liquid model 
(see Appendix~\ref{LL_Appendix}),  the phonon mode couples only to the charge degrees of freedom. 
At weak couplings, $\lambda=0.15$, the dynamical charge structure factor in Fig.~\ref{Dyn_ch.fig} 
shows precisely this feature; the continuum of charge excitations is supplemented by the 
dispersionless phonon mode at $\omega_0 = 0.1t$. In the spin sector (see Fig.~\ref{Dyn_sp.fig}) 
only the continuum of two spinon excitations is present.  

At larger values of $\lambda$  ($\lambda = 0.35$)  and as a consequence of the bipolaron formation 
spectral weight at low energies in the dynamical spin structure factor 
is suppressed.  In particular from  Fig. \ref{Dyn_sp.fig}  we can obtain  a rough estimate of the  
spin gap at, $\Delta_{sp} \simeq 0.2 t $ at $\lambda = 0.35$.
The lattice distortion in the Peierls  phase is accompanied by a softening of the phonon mode. 
At   $\lambda = 0.35$ (see Fig. \ref{Dyn_ch.fig})  we  observe a piling up of 
spectral weight  at  very low frequencies with dominant spectral intensity at 
${\pmb q} = 2k_F $.  This low energy feature  corresponds 
to the slow charge dynamics of the bipolaronic  $2k_f$ CDW 
(see Fig. \ref{Static.fig}a) \footnote{This slow dynamics of  the bipolarons 
is  at the origin of long autocorrelation times  observed in the QMC  simulations at 
large values of $\lambda$.}.  The high-energy continuum at  
$\lambda = 0.35$ in  $N({\pmb q} , \omega)$ is comparable to  $S( {\pmb q}, \omega) $  at the 
same coupling.  This  similarity confirms  that this structure stems 
from the particle-hole bubble of dressed  single particle Green functions.

We note that phonon dynamics have been studied for the spinless Holstein model within 
a projector based  renormalization method \cite{Sykora06} as well as with exact diagonalization 
and CPT methods \cite{Hohenadler06}.  In analogy to our results,  the phonon spectral 
function reveals not only the phonon dynamics but also the particle-hole continuum.  

\begin{figure}[ht]
\includegraphics*[width=0.35\textwidth]{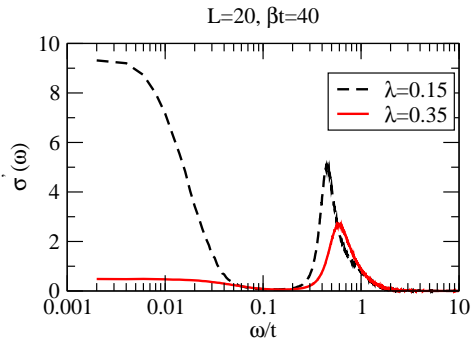} 
\caption{Optical conductivity  in the Luttinger liquid phase and bipolaronic CDW phases.
The calculations were carried out with periodic boundary conditions. For this choice of 
boundary conditions, the sum rule $\int {\rm d}  \omega \sigma'(\omega) = 
-\pi \langle \hat{K} \rangle $ 
where $\hat{K}$ is the kinetic energy, holds only in the thermodynamic limit.  In the plot, 
we have imposed this sum rule by normalizing the spectra by an overall factor.  } 
\label{Cond.fig}
\end{figure}

Finally we consider the real part of the optical conductivity,
\begin{equation}
    \sigma'(\omega) = 
    \frac{\pi}{Z \omega } \sum_{n,m} e^{-\beta E_m } (1-e^{-\beta \omega}) | \langle n 
    | \hat{j} | m \rangle|^2 
\delta( E_n -  E_m - \omega )   
\end{equation}
with $\hat{j} = i t \sum_{ {\pmb  i},\sigma } \left( 
\hat{c}^{\dagger}_{ {\pmb i},\sigma}  \hat{c}^{}_{ {\pmb i} + {\pmb a},\sigma } - {\rm H.c.  }
\right) $  both  at $\lambda=0.15$ and $\lambda = 0.35$.   Our results on an $L=20$ lattice are 
plotted in Fig. \ref{Cond.fig}.  As apparent at $\lambda = 0.15$ a  Drude feature reflecting  polaronic conductivity is  visible.  In contrast, at larger electron-phonon couplings, 
the  formation of the bipolaronic CDW  leads to  a substantial  suppression of the 
Drude feature.  The suppression  of the Drude weight reflects  the  very small charge velocity 
of the bipolarons. This follows from  the continuity equation which establishes a 
relation between the  optical conductivity and the  dynamical charge structure factor: 
\begin{equation}
\label{SigmavsNq.eq}
	\sigma'( {\pmb q},\omega)   = \frac{\omega}{ {\pmb q}^2 } 
        \left( 1 - e^{-\beta \omega} \right) N( {\pmb q},\omega).
\end{equation}
At small  momentum transfer, and using the sum rule 
$\frac{1}{\pi} \int {\rm d} \omega  N({\pmb q}, \omega) = N( {\pmb q} )$, 
we can model the dynamical charge structure factor by: 
$N({\pmb q},\omega)  = \pi N({\pmb q}) \delta(v_c {\pmb q} - \omega)  $ 
with $v_c$ the charge velocity.  
From Fig. \ref{Static.fig}a 
$N({\pmb q}) \propto  {\pmb q} $ in the long wavelength limit, and the 
 proportionality constant  is  to a good approximation $\lambda $ independent.  Inserting this 
approximate form of  into Eq. (\ref{SigmavsNq.eq})  gives in the zero temperature limit:
\begin{equation}
	\lim_{{\pmb q} \rightarrow 0} \sigma'( {\pmb q}, \omega )    
\propto  v_c \delta(\omega).
\end{equation} 
Hence, the suppression of the Drude weight stems from reduction of the charge velocity when 
passing from the Luttinger liquid phase to the bipolaronic CDW phase.

\subsection { Temperature dependence of the single particle spectral function} 
\label{Akom}
In this section we study the details of the temperature dependence of the single particle 
spectral function, both in the Luttinger and bipolaronic CDW phases. 

\subsubsection{Atomic limit}
\begin{figure}[ht]
\includegraphics*[width=0.35\textwidth]{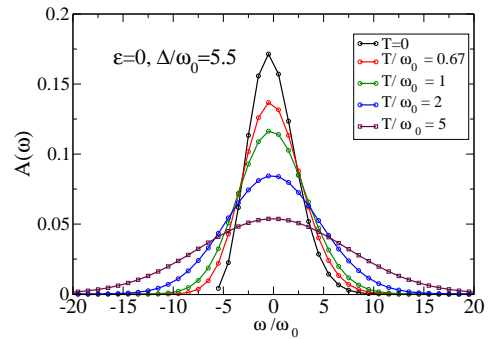} 
\caption{Spectral function  as a function of temperature in the atomic limit. } 
\label{Atomic.fig}
\end{figure}

It is instructive to start with  the atomic limit, $t = 0$, and in the 
absence of spin degrees of freedom,   
\begin{equation}
   	 \hat{H} = \epsilon \hat{c}^{\dagger} \hat{c}  
           + g  \hat{Q} \hat{n}   
                + \frac{\hat{P}^2}{2M} + \frac{k}{2} \hat{Q}^2,
\end{equation}
where exact solutions for the temperature dependence of the spectral function are available 
\cite{Mahan90}. 
In particular, at $T=0$, (see Fig. \ref{Atomic.fig}) 
the single particle  spectral function is given by: 
\begin{equation}
	A(\omega) =  e^{-\Delta/\omega_0}\sum_{l =0}^{\infty} \frac{1}{l!} 
 \left( \Delta / \omega_0 \right)^l 
 \delta( \omega - \left[ \epsilon - \Delta + \omega_0 l \right] ) 
\end{equation}
with $\Delta = g^2/2k$.  

An electron on the energy level couples to the phonon degrees of
freedom and can lower its energy at the expense  of a  shift in the ground state expectation 
value of $ \hat{Q} $. Thereby $ \epsilon \rightarrow \epsilon - \Delta$  which 
corresponds to the lowest energy pole in $ A(\omega) $.  Since the ground state contains an 
infinite  number of phonon excitations, poles at $ \epsilon - \Delta + \omega_0 l$ following
a Poisson distribution are apparent in the single particle spectral function.   The spectral 
function is centered around  
$ \langle \omega \rangle \equiv \int {\rm d } \omega A(\omega) w   = \epsilon $   and 
has a width $ \sqrt{ \langle \left[ \omega - \langle \omega \rangle \right]^2 \rangle   } = 
\sqrt{\omega_0 \Delta} $.  The relevant energy scale for  the temperature  behavior of the 
spectral function is the  phonon frequency, $\omega_0$.  As apparent in Fig. \ref{Atomic.fig} 
at temperatures in the vicinity  of the phonon frequency  a considerable  broadening  of the 
spectral function is apparent. 

\subsubsection{ Luttinger Liquid phase, $\lambda = 0.25 $ }

\begin{figure}[ht]
\includegraphics[width=3.5cm,height=6cm]{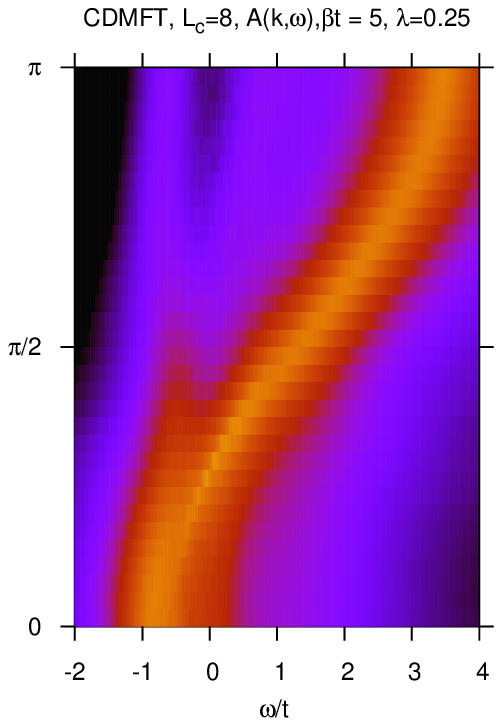} 
\includegraphics[width=4.5cm,height=6cm]{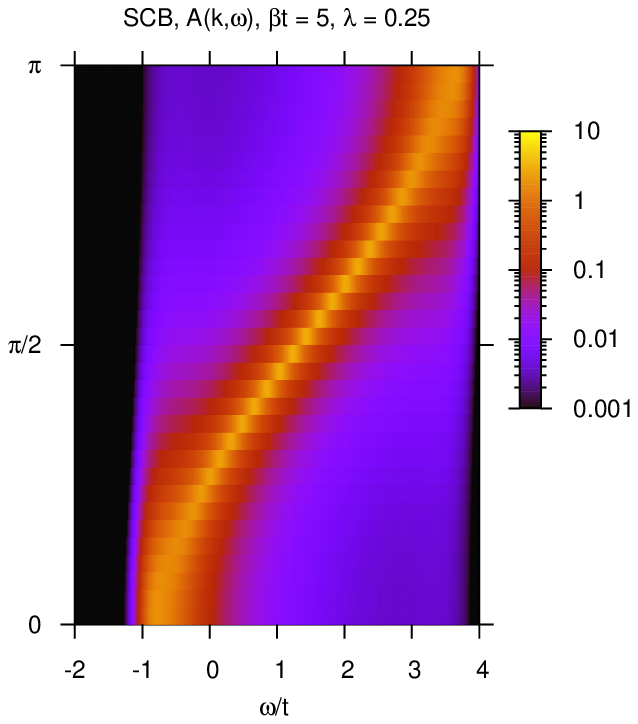}   \\
\includegraphics[width=3.5cm,height=6cm]{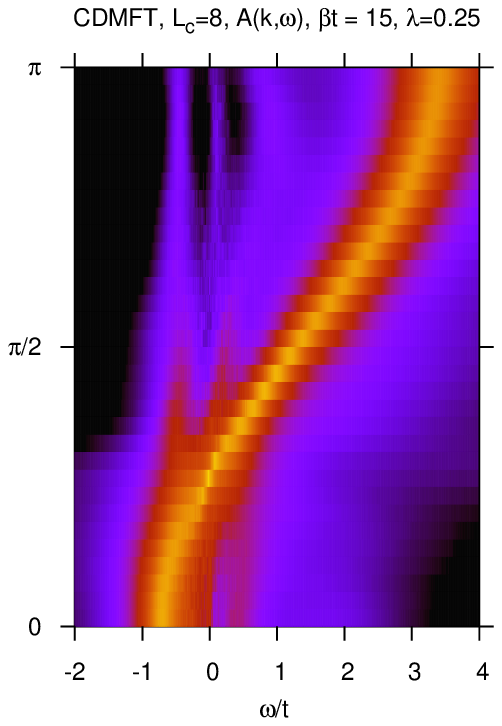}
\includegraphics[width=4.5cm,height=6cm]{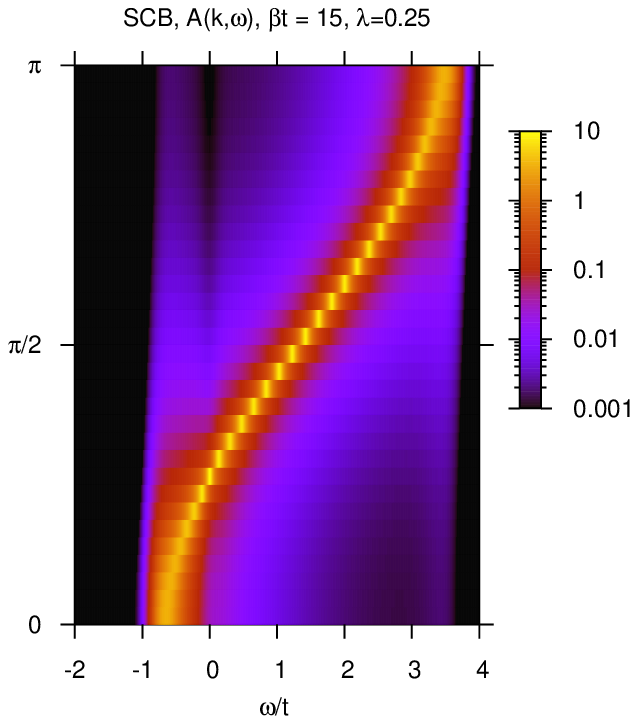}  \\
\includegraphics[width=3.5cm,height=6cm]{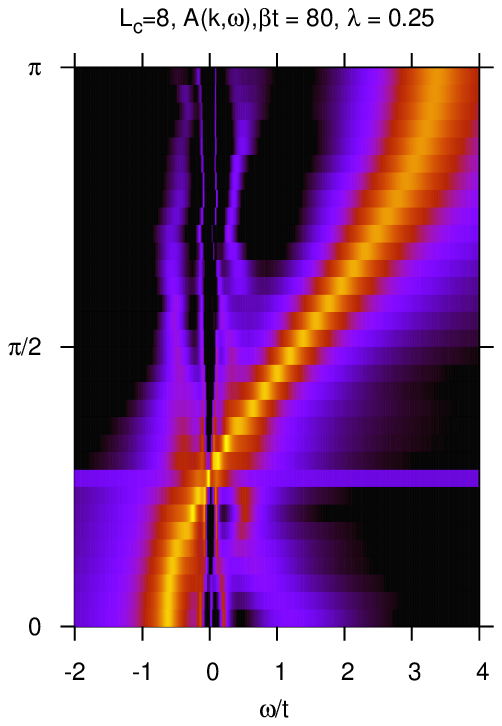}
\includegraphics[width=4.5cm,height=6cm]{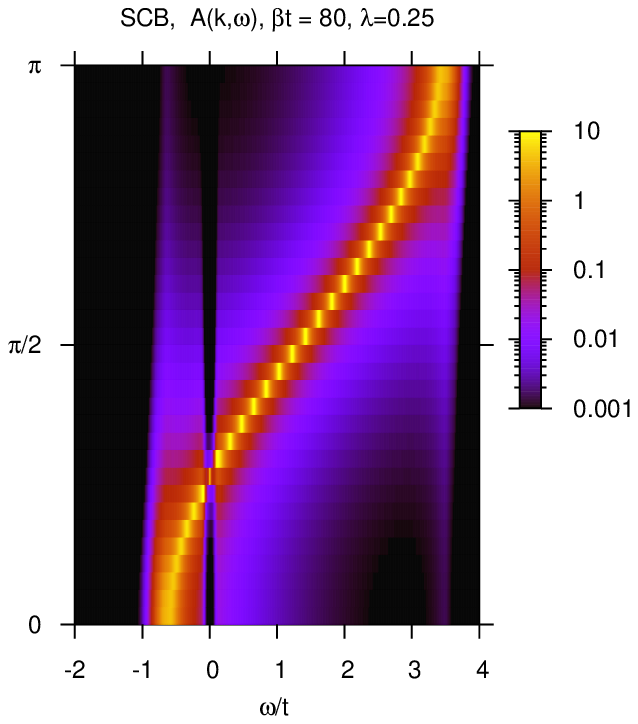} 
\caption{ Single particle spectral function at various temperatures. 
The left hand panels are CDMFT calculations on $L_c=8$ clusters. The right hand 
panels provide a comparison with the SCB approximation (See Appendix \ref{SCB.SEC}) 
The y-axis corresponds to the  crystal momentum $\pmb k $.  } 
\label{LL_Akom.fig}
\end{figure}

Fig. \ref{LL_Akom.fig} plots the temperature dependence of the single 
particle spectral function for the Holstein model in the Luttinger liquid phase at 
$\lambda = 0.25$.  We compare our results to the self-consistent Born (SCB) 
approximation \cite{Engelsberg63} briefly reviewed in Appendix \ref{SCB.SEC}.  
At  high temperatures, $T/\omega_0 \geq 1$, 
the overall features of the spectral  function as obtained from the SCB compare favorably with 
the  CDMFT  calculations. Both  show a broad  spectral function  centered around  the bare 
electron energy $ \epsilon(k) - \epsilon(k_F) $.  As in the atomic limit and at an energy scale 
set by the phonon frequency  a  substantial narrowing of the spectral function and reordering of 
spectral  weight is apparent.  

\begin{figure}
\begin{center}
\includegraphics*[width=0.25\textwidth]{LLiquid.eps}  \\
\includegraphics[width=0.25\textwidth]{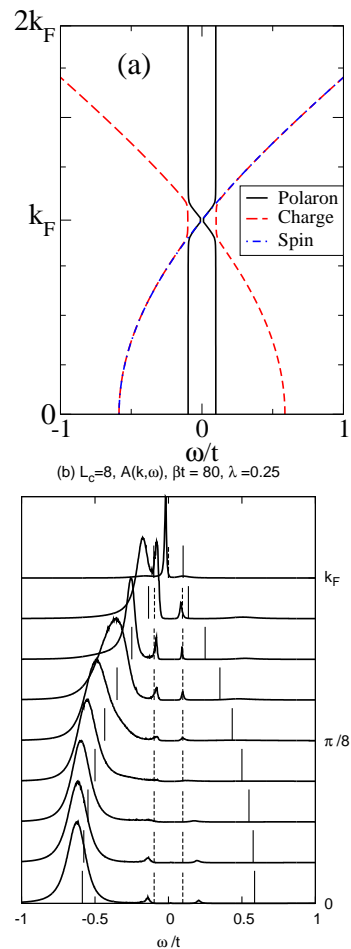} 
\end{center}
\caption{ Dispersion relation of spin and mixed phonon and charge modes as obtained from the 
the Luttinger liquid forward scattering Hamiltonian of Eq. \cite{Meden94}. 
Here we have set $\lambda  = 0.25$, 
$\omega_0  = 0.1t$ and $ k_F = \pi/4 $ as appropriate for quarter filling. For the purposes 
of comparison with the QMC data, we have taken the liberty of replacing 
$v_F {\pmb k} $ by $ -2t \cos({\pmb k}a) + 2t \cos({\pmb k}_F a) $ } 
\label{LLiquid.fig}
\end{figure}

As the temperature  drops well below the phonon frequency, $\beta t = 80$,  the CDMFT spectral 
function  exhibits sharp  features which are not captured by the SCB approximation.   For 
instance at $\omega /t > 0$ and  $ k < k_F $ a sharp peak is apparent at $ \omega \simeq \omega_0 $
in the QMC spectra  and is not  present in the SCB approximation.    Of course, the SCB 
approximation has many caveats since i) it does not contain vertex corrections  required in the 
low-temperature Luttinger liquid phase  and ii) the phonon  propagator is 
not renormalized such that phonon softening and signatures of the Peierls transitions are 
not included in the  approximation.  The low temperature CDMFT spectral function  at 
$\lambda = 0.25$  is at best understood within the framework of bosonization as sketched 
in Appendix~\ref{LL_Appendix}. In a first approximation, and deep in the Luttinger liquid phase, 
one can neglect backward scattering \cite{Voit86} thereby obtaining  the forward 
scattering model of Eq. (\ref{H_LL}) \cite{Meden94} containing  spin, phonon and charge modes.
The spin mode decouples and the charge and phonon mode mix.
At the expense of a  Bogoliubov transformation, the forward scattering model can be 
diagonalized to obtain the  dispersion relations   shown in Fig.~\ref{LLiquid.fig}a. 
Gapless spin and polaron modes as well as  a gapfull charge mode are apparent.  
Since  the single  electron operator 
can be expressed in terms of the spin and charge  operators \cite{Giamarchi}  one expects signatures
of those modes in the  single particle spectral function.  Fig.~\ref{LLiquid.fig}b  plots a 
closeup of  $ A({\pmb k}, \omega)$, at our lowest temperature. Structures  following the  coupled 
gaped charge and polaron modes (vertical lines)  are clearly apparent.   
According to Fig. \ref{LLiquid.fig}a
the spin mode is next to degenerate with the charge modes, and hence difficult to detect in 
our numerical calculations. 
	
\subsubsection{ Peierls phase, $\lambda = 0.35 $ }
\begin{figure}[h]
\includegraphics[width=3.5cm,height=5.5cm]{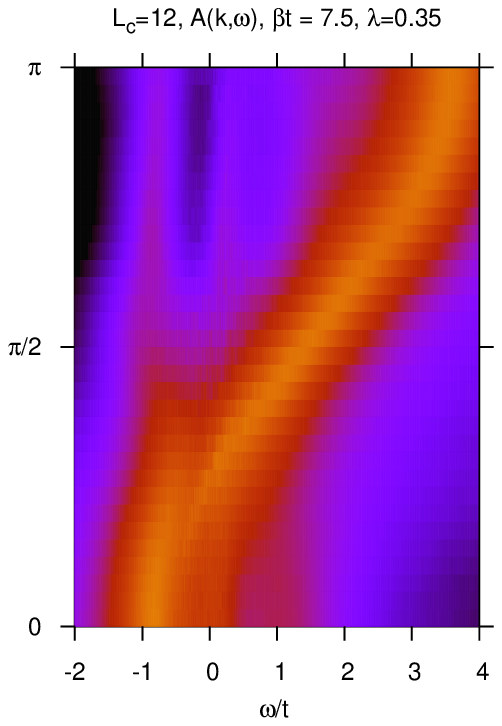}  
\includegraphics[width=3.5cm,height=5.5cm]{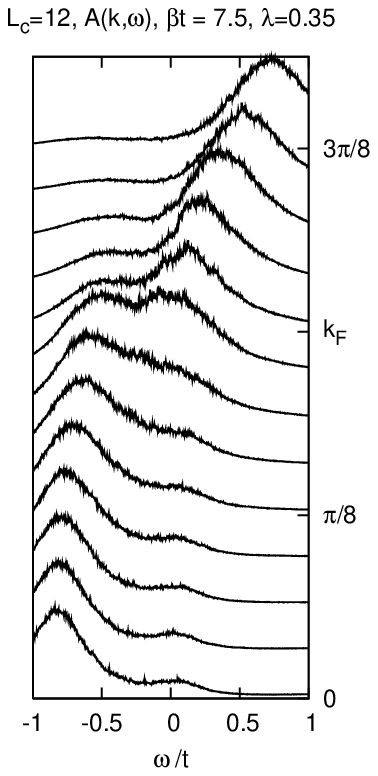}    \\
\includegraphics[width=3.5cm,height=5.5cm]{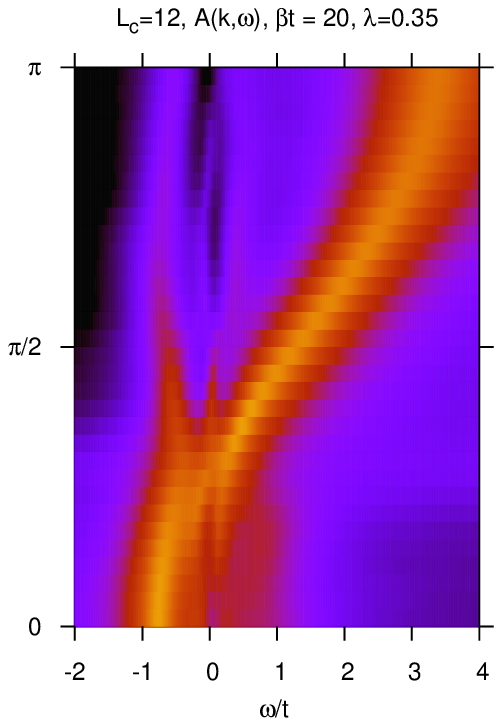}   
\includegraphics[width=3.5cm,height=5.5cm]{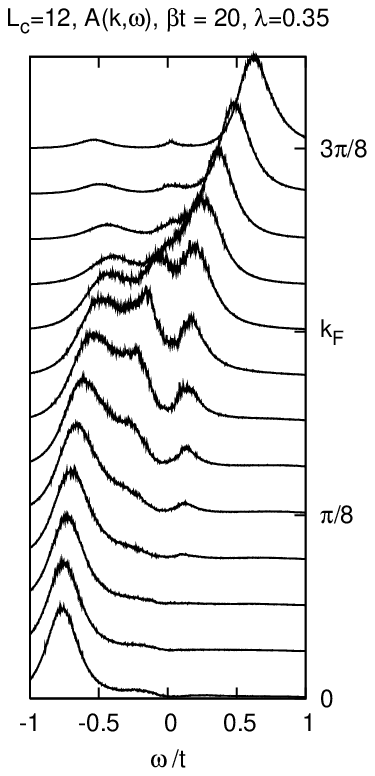}   \\
\includegraphics[width=3.5cm,height=5.5cm]{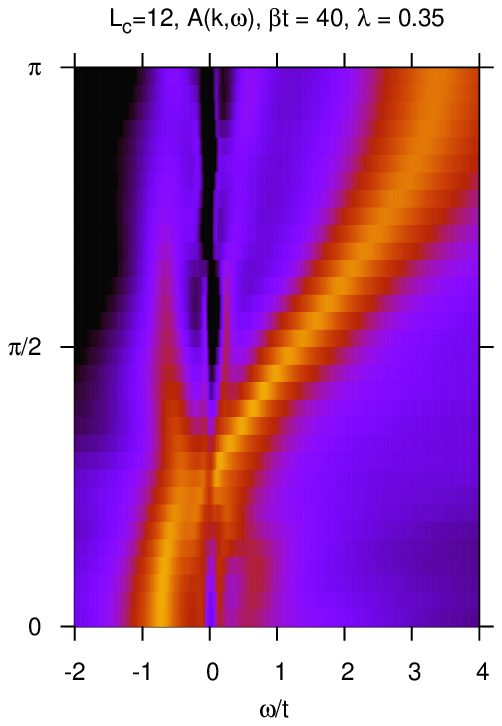}   
\includegraphics[width=3.5cm,height=5.5cm]{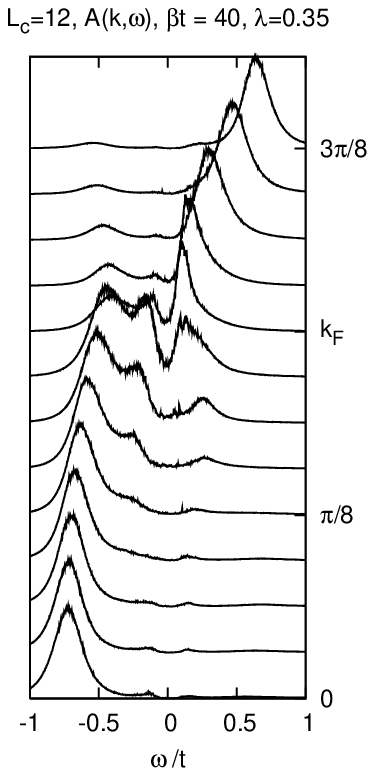}   
\caption{Temperature dependence of the single particle spectral function  in the Peierls phase. 
The results stem from CDMFT on an $L_c=12$ cluster. The left panels correspond to logarithmic 
intensity plots with scale given in  Fig. \ref{LL_Akom.fig}. The right panels shown the 
spectral function in a narrow window around the Fermi energy and momentum. Here, we have normalized 
the maximal peak height to unity, and the total weight under the spectral function is 
given by: $\int_{-\infty}^{\infty} {\rm d} \omega A({\pmb k},\omega) = \pi$.
\label{Peierls_Akom.fig} }
\end{figure}
At larger values of the electron-phonon coupling backward-scattering becomes relevant and is at 
the origin of the Peierls transition.   
Fig. \ref{Peierls_Akom.fig} tracks the 
temperature dependence of the  single particle  spectral function  at $\lambda = 0.35$  
which places us in the Peierls phase.
At high temperatures, $ 1/\beta > \omega_0 $ the overall features can  again  be well accounted 
for within the SCB approximation reviewed in Appendix \ref{SCB.SEC}. 
Upon cooling 
(see the $\beta t = 20$ data set in Fig. \ref{Peierls_Akom.fig}) a narrow polaronic band  crosses 
the Fermi  energy, and  gaped higher energy 
excitations show precursor features of back-folding. 
This data set shows remarkable similarities  with the features  observed in the Luttinger liquid 
phase thereby suggesting that  aspects of the Luttinger liquid spectral functions are 
apparent at finite temperatures above the crossover to the Peierls phase.  
At our lowest temperature, the 
narrow polaronic band develops a gap of the order $2 \Delta_{qp} \simeq 0.2t$, giving rise to 
rather 
dispersionless features  in the spectral function at $\omega  \simeq  0.1 t$.    
We interpret those features  in terms of the formation of the bipolaronic CDW. 
Here, removing an electron costs the bipolaron binding energy. The fact the that the 
spin gap  at $\lambda = 0.35$ as obtained from Fig. \ref{Dyn_sp.fig} matches  $2 \Delta_{qp} $ 
confirms this interpretation.

\subsection{Interpretation in terms a transition from a Luttinger  to 
a Luther-Emery liquid}
A very natural account of the above presented  data stems from  a transition between    
Luttinger  and Luther-Emery liquids.  The Luther-Emery liquid description of the 
Peierls phase has  been put forward  by Voit \cite{Voit98}. 
Within this framework and away from half filling  umklapp processes leading to a charge gap are  
absent. Note however that  at quarter  band filling,
second order  umklapp processes are allowed and will lead to a charge gap provided that 
the interactions are strong enough such that  $K_\rho  < 1/4$ \cite{Giamarchi97}. 
Here we omit this possibility  since it does not naturally  explain our numerical data on 
small lattices  and $\lambda \leq 0.35$. 
Backward scattering  on the other hand is present and  if relevant can lead to the opening of 
a spin gap leaving the charge sector gapless.  This corresponds to the Luther-Emery liquid. 

The Luttinger liquid fix-point is characterized  by 
dominant forward scattering processes and the asymptotic behavior of correlation functions 
is governed by single dimensionless quantity, $K_{\rho}$. 
Neglecting logarithmic  corrections \cite{Schulz90}  the correlation functions read:
\begin{equation}
\label{LL.eq}
\begin{gathered}
 \langle n({\pmb r}) n({\pmb 0}) \rangle = 
 \frac{K_{\rho}}{\left(\pi {\pmb r}\right)^2 } + A_1 \cos(2{\pmb k}_f {\pmb r} )
  {\pmb r}^{-1-K_\rho} + \cdots \\ 
    +  A_2 \cos(4{\pmb k}_f {\pmb r} )   {\pmb r}^{-4 K_{\rho} }  \\
 \langle {\pmb S} ({\pmb r}) {\pmb S}({\pmb 0}) \rangle =  
\frac{1}{\left( \pi {\pmb r} \right)^2} + B_1 \cos(2{\pmb k}_f {\pmb r} ) 
{\pmb r}^{-1-K_\rho} + \cdots \\
 \langle \Delta^{\dagger}({\pmb r}) \Delta({\pmb 0}) \rangle =  
C {\pmb r}^{-1-1/K_\rho} + \cdots
\end{gathered}
\end{equation}
Logarithmic corrections do not  show up in the first term of the charge-charge correlation 
functions \cite{Schulz90} and hence allow an efficient determination of $K_{\rho}$  via: 
\begin{equation}
      K_{\rho}  =  \pi \lim_{{\pmb q} \rightarrow  0} \frac{  {\rm d} N( {\pmb q}) }
                                                           {  {\rm d}  {\pmb q} }.
\end{equation}
Form our data  on an admittedly small lattice, $L=20$, we obtain from the above equation:
\begin{equation} 
\begin{gathered}
K_{\rho} = 1.0341 \pm 0.0006  \; \;  {\rm at }  \; \; \lambda = 0.15 \\
K_{\rho} = 1.0441 \pm 0.0002  \; \;  {\rm at }  \; \; \lambda = 0.25
\end{gathered}
\end{equation}
Since $K_{\rho}  > 1$  one would conclude  that the Luttinger liquid phase is characterized 
by dominant superconducting correlations. 

The Luther-Emery liquid has correlation functions which read: 
\begin{equation}
\label{LE.eq}
\begin{gathered}
 \langle n({\pmb r}) n({\pmb 0}) \rangle = 
 \frac{A_0}{ {\pmb r}^2 } + A_1 \cos(2{\pmb k}_f {\pmb r} )
  {\pmb r}^{-K_\rho} + \cdots \\ 
    +  A_2 \cos(4{\pmb k}_f {\pmb r} )   {\pmb r}^{-4 K_{\rho} }  \\
 \langle \Delta^{\dagger}({\pmb r}) \Delta({\pmb 0}) \rangle =  
C {\pmb r}^{-1/K_\rho} + \cdots
\end{gathered}
\end{equation}
and an exponential decay of the spin-spin correlations \cite{Troyer93}. 
Assuming the validity of the above, we can deduce a rough  estimate 
of the value of $K_\rho$ in the Luther-Emery phase.
Since our data at $\lambda = 0.35$ shows dominant $2k_f$ charge fluctuations, 
we conclude that $K_\rho  < 1$ in the Peierls phase. 
A more precise upper bound for $K_\rho$  can be obtained by comparing the pairing 
correlation functions at $\lambda = 0.25$ in the Luttinger liquid phase and at $\lambda = 0.35$. 
At $\lambda = 0.25$, $K_\rho$ is slightly larger that unity such that the pairing correlations 
fall of as $r^{-1.958}$. As apparent from Fig. \ref{Static.fig}b, the pairing correlations at 
$\lambda =   0.35$  in the Luther-Emery phase fall off quicker, thus implying  $K_\rho < 1/2$ 
in the Luther-Emery phase at $\lambda = 0.35$. 
This upper bound,  $K_\rho < 1/2$,  equally implies  a sub-dominant $4k_f$ 
charge density decaying  more slowly than $r^{-2}$. The observed $4k_f$ cusp in the  static charge 
structure factor at $4k_f$ and $\lambda =0.35$   (see Fig. \ref{Static.fig}a) is consistent 
with this remark.

\section{Conclusions}

In conclusions we have  used a generalization of the diagrammatic determinantal QMC  
algorithm, to investigate the physics of the quarter-filled  one-dimensional Holstein model. 
We have used the algorithm for lattice simulations to extract two particle 
quantities in the context of CDMFT to investigate the temperature dependence of the single 
particle spectral function both in the Peierls and Luttinger liquid phases.  

Our results are naturally interpreted in terms of a transition from  Luttinger  to Luther-Emery 
liquids.  The Luttinger liquid phase has a $K_\rho$ which is marginally greater than unity such that 
pairing correlations are dominant. At  our considered phonon frequency, $\omega_0/t = 0.1$, the 
Luther-Emery phase is characterized  by $K_\rho < 1/2$ and thereby by dominant $2k_f$ charge 
fluctuations. At even large  values of $\lambda$ than considered in this article,  
one can  expect $K_\rho$ to drop below the  $1/4 $ threshold triggering the  opening a gap also in the 
charge sector via second order umklapp.  
Hence  at this commensurate  filling and adiabatic phonon frequency, we can speculate  the phase 
diagram  as a function of $\lambda$ to not only show transition  between Luttinger and Luther-Emery 
liquids but also at $\lambda > 0.35$  a transition from the Luther-Emery phase to 
a fully gaped phase both in the charge and spin sectors.   One equally expects the character of the 
Luther-Emery phase to very dependent on the phonon frequency. In the antiadiabatic limit  
the Holstein model maps onto the attractive Hubbard model where superconducting correlations 
are dominant such that $K_\rho > 1$. 

Our calculations equally reveal the rich temperature dependence of the single particle spectral 
functions. We can access a  temperature range  covering the 
domain of validity of the self-consistent Born  approximation in the high temperature limit down to 
to temperatures  where the  Luttinger liquid or Luther-Emery fix points are relevant.  The 
temperature dependence in the Luther-Emery phase interestingly shows that above the temperature 
scale at which the single gap opens at the Fermi energy, features of the Luttinger liquid phase, 
namely a polaronic band  crossing the Fermi energy and a gaped charge mode, are apparent. 
This observation  should be set in the context of photoemission  experiments carried out on 
TTF-TCNQ organics  where measurements are carried  out at a  temperature scale above 
the Peierls transition and 
interpreted in terms of a Luttinger liquid model 
\cite{Claessen03,Jeckelmann04,Abendschein06,Bulut06}.

\acknowledgements I would like to thank  H. Fehske, N. Nagaosa, T. Lang  and S. Capponi 
for  comments and discussions.  
The  simulations were carried out on the IBM p690 
at the John von Neumann Institute for Computing, J\"ulich. I would like to thank this institution
for generous allocation of CPU time. 
Financial support from the DFG under the grant number AS120/4-2 and the DAAD  in terms of a PROCOPE
exchange program is acknowledge.  

\appendix 
\section{ Self-consistent Born approximation }
\label{SCB.SEC}
For the Holstein model given by Eq. (\ref{Holstein_Ham}), 
the self-energy diagram shown in  Fig. \ref{SCB.fig}  can be evaluated to give:
\begin{figure}
\begin{center}
\includegraphics[width=0.2\textwidth]{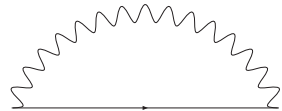} 
\end{center}
\caption{ Self-energy diagrams included in the self-consistent Born approximation.   
The solid (wavy) lines corresponds to the bare single particle 
Green function (phonon propagator) } 
\label{SCB.fig}
\end{figure}
\begin{eqnarray}
\label{Sigma_1}
	\Sigma_1(i \omega_m) = & & \frac{g^2 }{2k} {\omega_0} \frac{1}{L} \sum_{\pmb k} 
\left\{   \frac{n_B(\omega_0) + 1 - f[\epsilon(\pmb k)] } { - \epsilon(k) - \omega_0 + i \omega_m} 
 \right.  \nonumber \\
& & + \left.  \frac{n_B(\omega_0) + f[\epsilon(\pmb k)] } { - \epsilon(k) + \omega_0 + i \omega_m} 
 \right\}
\end{eqnarray}
Here,  $ f[\epsilon(\pmb k)] = \frac{1}{e^{\beta\epsilon(\pmb k)} + 1}$ is the Fermi function (note that we have 
included the chemical potential in the very definition of $\epsilon({\pmb k})$), 
 $ n_B(\omega_0) = \frac{1}{e^{\beta \omega_0} - 1}$ the Bose-Einstein distribution and 
$\omega_0 = \sqrt{\frac{k}{M}}$  the phonon frequency.  At zero temperature and for real frequencies, 
the imaginary part of the self-energy takes the form: 
\begin{eqnarray}
\label{ImSigma_1}
	{\rm Im} \Sigma_1(\omega) = & & 
- \frac{g^2 }{2k} {\omega_0} \frac{\pi}{L} \sum_{\pmb k} 
\left\{ \Theta[\epsilon(\pmb k) ] \delta(- \epsilon(k) - \omega_0 + \omega)   \right. \nonumber \\
& &+ \left. \Theta[-\epsilon(\pmb k) ] \delta(- \epsilon(k) + \omega_0 + \omega) \right\}.
\end{eqnarray}
The first (second) term  in Eq. (\ref{ImSigma_1}) corresponds to absorption (emission) 
of a phonon. Energy conservation as well as phase space limit those processes to energy range 
$\omega >  \omega_0$ for absorption and  $\omega  < -\omega_0 $  for emission.   Hence at $T=0$ 
and  in a region  of width 
$ 2 \omega_0$ centered  around the Fermi energy,  the imaginary part of the self-energy vanishes. 
In this range the single particle Green function has poles  defining a dispersion relation with 
effective mass: 
\begin{equation}
\label{Meff}
	\frac{m*}{m} = \left[1 - \frac{\partial {\rm Re} \Sigma(\omega) }{\partial \omega} \right]^{-1}_{\omega = 0}
\end{equation}

To obtain a good agreement  with the high temperature Quantum Monte Carlo data  
we sum up the  non-crossing self-energy diagrams.  This amounts  
to solving the set of self-consistent equations: 
\begin{eqnarray}
\label{SCB}
	G({\pmb k}, i \omega_m) & = & 
        \frac{1} { G_{0}^{-1}({\pmb k}, i \omega_m)  - \Sigma( i \omega_m ) } 
\\
        \Sigma( i \omega_m ) & = & \frac{g^{2} \omega_0}{2k} \frac{1}{\beta L} \sum_{ { \pmb k }, i \Omega_m}    
	D(i  \Omega_m) G( {\pmb k}, i \omega_m - i \Omega_m).
\nonumber 
\end{eqnarray}
Here, $D(i  \Omega_m) = \frac{1}{\omega_0 + i\Omega_m}  + \frac{1}{\omega_0 - i\Omega_m}$ is the bare phonon propagator
and $ \Omega_m $ a bosonic Matsubara frequency. 
Since at a given iteration  we do not have at hand the pole structure 
of $G( {\pmb k}, i \omega_m)$ in the complex frequency plane,  it is more convenient to solve the 
above equations  numerically for real frequencies. To do so, we use the  spectral 
representation of the Green function: 
\begin{equation}
	G({\pmb k}, i \omega_m) = \int {\rm d} \omega' \frac{ A(k,\omega') }
	                   { i \omega_m - i \Omega_m - \omega' }
\end{equation}
where $ A(k,\omega') = - \frac{1}{\pi} G^{{\rm ret}}({\pmb k},  \omega' ) $. 
With this choice and  $N(\omega') \equiv \frac{1}{L} \sum_{ {\pmb k} } A(k,\omega')$ the 
self-energy  reads: 
\begin{eqnarray}
	\Sigma( i \omega_m ) & = & \frac{g^2}{2k} \omega_0 
     \int{\rm d} \omega' N(\omega') \left\{ 
\frac{n_B(\omega_0) + 1 - f[\omega'] } { - \omega' - \omega_0 + i \omega_m} 
 \right.  \nonumber \\
& & + \left.  \frac{n_B(\omega_0) + f[\omega'] } { - \omega' + \omega_0 + i \omega_m} 
 \right\}.
\end{eqnarray}
At a given iteration step at which $N(\omega)$ is known we can compute with the above equation 
the self-energy on the real frequency axis ($i\omega_m \rightarrow \omega + i\delta $) and thereby 
recompute the single particle Green function and corresponding $N(\omega)$. Typically, for the 
considered parameter range, ten iterations suffice to achieve convergence. 

This approximation has many caveats.  Since the phonon propagator is not renormalized, phonon 
softening and hence  the Peierls transition is absent.  At low temperatures,  in the metallic 
phase, one equally expects the approximation to fail since it does not contain vertex corrections 
necessary to produce the Luttinger liquid physics.    

\section{Luttinger-Liquid}
\label{LL_Appendix}
At low temperatures the self-consistent Born approximation does not capture the expected Luttinger 
behavior of the one-dimensional Holstein model.  Neglecting backscattering -- an approximation which
one can justify in the Luttinger liquid phase --  exact solutions  at asymptotically  low
energy scales  are possible.  Here, we briefly outline the steps. 
	With the bosonic raising and lowering operators 
\begin{equation}
	\hat{a}_{\pmb i} =  \frac{\omega_0 M \hat{Q}_{\pmb i} +  i \hat{P}_{\pmb i} }
                            { \sqrt{2 \omega_0 M } }
\end{equation}
satisfying the bosonic commutation rules 
$ \left[ \hat{a}_{\pmb i}, \hat{a}^{\dagger}_{\pmb j} \right] = \delta_{{\pmb i}, {\pmb j} } $,
the Holstein model reads: 
\begin{eqnarray}
  \hat{H}  = & & \sum_{ \pmb{k},\sigma} \epsilon({\pmb k}) \hat{c}^{\dagger}_{ {\pmb k},\sigma}
                       \hat{c}_{{\pmb k}, \sigma} + 
                    \omega_0\sum_{\pmb q} \hat{a}^{\dagger}_{\pmb q} \hat{a}_{\pmb q} + \nonumber \\
	   & & \frac{ g }{ \sqrt{ 2 \omega_0 M } }  \frac{ 1 }{ \sqrt{L} } \sum_{q} 
         \hat{c}^{\dagger}_{{\pmb k},\sigma}  \hat{c}_{\pmb k +\pmb q, \sigma} 
        \left(  \hat{a}^{\dagger}_{\pmb q} + \hat{a}_{-{\pmb q} } \right) 
\end{eqnarray}
where the Fourier transform is defined as,
\begin{equation} 
	 \hat{c}^{\dagger}_{{\pmb k},\sigma} = \frac{1}{\sqrt{L}} \sum_{\pmb i} 
        e^{i {\pmb k} {\pmb i} } \hat{c}^{\dagger}_{ 	{\pmb i},\sigma}
\end{equation}
with an equivalent definition for the bosonic phonon operators ${\hat a}_{\pmb q}$.

Linearization around the Fermi points and introducing left ($\hat{L}_{\pmb k,\sigma}$) and 
right ($\hat{R}_{\pmb k,\sigma} $ ) fermionic creation  operators yields the effective low
energy form for the kinetic energy term,
\begin{equation}
	 \sum_{ \pmb{k}, \sigma } \epsilon({\pmb k}) \hat{c}^{\dagger}_{ {\pmb k},\sigma}
                           \hat{c}_{ {\pmb k} , \sigma } \rightarrow 
	\sum_{ \pmb{k}, \sigma } v_F {\pmb k} \left(  
\hat{R}^{\dagger}_{\pmb k,\sigma} \hat{R}_{\pmb k,\sigma} 
-\hat{L}^{\dagger}_{\pmb k,\sigma} \hat{L}_{\pmb k,\sigma}  \right)  
\end{equation}
which in its bosonized form  reduces to: 
\begin{eqnarray}
	  & &  \sum_{ {\pmb q},\sigma } v_F|{\pmb  q}| \hat{b}^{\dagger}_{ {\pmb q}, \sigma} 
\hat{b}_{ {\pmb q}, \sigma}  \; \; \;  {\rm with}  \nonumber \\ 
	\hat{b}_{{\pmb q}, \sigma}  & & = 
\left\{
\begin{array}{cc}
\left(\frac{2 \pi}{|q|L} \right)^{1/2}\sum_{\pmb k} \hat{R}^{\dagger}_{\pmb k,\sigma} 
  \hat{R}_{{\pmb k} + {\pmb q} ,\sigma} & {\pmb q } > 0 \\
  \left(\frac{2 \pi}{|q|L} \right)^{1/2}\sum_{\pmb k} \hat{L}^{\dagger}_{\pmb k,\sigma} 
  \hat{L}_{{\pmb k} + {\pmb q} ,\sigma} & {\pmb q } < 0 
\end{array}
\right. \\
{\rm and} \; \;
& &   \left[ \hat{b}_{ {\pmb q}, \sigma}, \hat{b}^{\dagger}_{ {\pmb q'}, \sigma'} \right] = 
    \delta_{ {\pmb q}, {\pmb q'} } \delta_{\sigma, \sigma'} 
\end{eqnarray}

	After linearization the electron-phonon interaction, in terms of left and right movers, 
reads: 
\begin{eqnarray}
	 \frac{g}{\sqrt{2 \omega_0 M L}} & & \sum_{{\pmb q}, {\pmb k}, \sigma} 
\left\{           \hat{L}^{\dagger}_{\pmb k,\sigma} \hat{R}_{{\pmb k} + {\pmb q}, \sigma} 
        \left( \hat{a}^{\dagger}_{{\pmb q}+2{\pmb k}_f} + \hat{a}_{-{\pmb q} -2 {\pmb k}_f } \right)
\right. \nonumber \\
+ & &
       \hat{R}^{\dagger}_{\pmb k,\sigma} \hat{L}_{{\pmb k} + {\pmb q}, \sigma} 
        \left( \hat{a}^{\dagger}_{{\pmb q}-2{\pmb k}_f} + \hat{a}_{-{\pmb q} + 2 {\pmb k}_f } \right)
 \\
+ & &
\left. \left(\hat{L}^{\dagger}_{\pmb k,\sigma} \hat{L}_{{\pmb k} + {\pmb q}, \sigma} +
         \hat{R}^{\dagger}_{\pmb k,\sigma} \hat{R}_{{\pmb k} + {\pmb q}, \sigma} \right)
        \left(  \hat{a}^{\dagger}_{\pmb q} + \hat{a}_{-{\pmb q} } \right)  \right\}.
\nonumber 
\end{eqnarray}
The first two terms correspond to back-scattering processes which lead to  enhanced $2k_f$ charge 
fluctuations, an enhanced effective mass and ultimately to the 
Peierls  phase.  To obtain a first description of the Luttinger liquid phase, 
we omit them thereby obtaining  a solvable model with only forward scattering  processes:
\begin{eqnarray}
\hat{H}_{LL}   & = &  \sum_{ {\pmb q},\sigma } v_F|{\pmb  q}| \hat{b}^{\dagger}_{ {\pmb q}, \sigma} 
\hat{b}_{ {\pmb q}, \sigma}  + \omega_0\sum_{\pmb q} \hat{a}^{\dagger}_{\pmb q} \hat{a}_{\pmb q} 
 \\
 & &  + \sqrt{\frac{g}{2 \omega_0 M \pi }} \frac{1}{\sqrt{2}} 
\sum_{{\pmb q},\sigma} |\pmb{q}| \left( \hat{b}^{\dagger}_{- { \pmb q}, \sigma} + 
\hat{b}_{{ \pmb q}, \sigma} \right) \left(  \hat{a}^{\dagger}_{\pmb q} + \hat{a}_{-{\pmb q} } \right)
\nonumber.
\end{eqnarray} 
With  spin and charge  densities defined as,
\begin{eqnarray}
	\hat{\sigma}_{\pmb q} & = & \frac{1}{\sqrt{2}} 
        \left(\hat{b}_{{\pmb q},\uparrow}  - \hat{b}_{{\pmb q},\downarrow} \right)  
\nonumber \\
	\hat{\rho}_{\pmb q} & = & \frac{1}{\sqrt{2}} 
        \left(\hat{b}_{{\pmb q},\uparrow}  + \hat{b}_{{\pmb q},\downarrow} \right),
\end{eqnarray} 
$\hat{H}_{LL} $ takes the form:
\begin{eqnarray}
\label{H_LL}
    \hat{H}_{LL} & = & 
\sum_{{\pmb q}} v_F|{\pmb  q}| \hat{\sigma}^{\dagger}_{ {\pmb q}} \hat{\sigma}_{ {\pmb q}} 
+ \sum_{{\pmb q}} v_F|{\pmb  q}| \hat{\rho}^{\dagger}_{ {\pmb q}} \hat{\rho}_{ {\pmb q}} 
+ \omega_0\sum_{\pmb q} \hat{a}^{\dagger}_{\pmb q} \hat{a}_{\pmb q} 
\nonumber \\
 +  & & \sqrt{\frac{g}{2 \omega_0 M \pi }} \sum_{{\pmb q} } |\pmb{q}|  
\left(  \hat{\rho}^{\dagger}_{-{\pmb q}} + \hat{\rho}_{{\pmb q}} \right) 
\left(  \hat{a}^{\dagger}_{\pmb q} + \hat{a}_{-{\pmb q} } \right)	
\end{eqnarray}
As apparent, the spin mode decouples and the charge and phonon modes mix. A  Bogoliubov transformation 
diagonalizes  the Hamiltonian and reveals the dispersion relation of those modes (see 
Fig. \ref{LLiquid.fig}).


\end{document}